\documentclass[12pt,preprint,,tightenlines]{revtex4}
 \usepackage{graphicx}
 \usepackage{dcolumn}
 \usepackage{amsmath}
 \usepackage{textcomp}

\begin{document}

 \title{Stark Field Modulated Microwave Detection of Molecular Chirality}
 
 \author{Kevin K. Lehmann}
 \affiliation{
 Departments of Chemistry \& Physics, University of Virginia, Charlottesville VA, 22904-4319}
 \email{lehmann@virginia.edu}
 \date{\today}

\begin{abstract}
Patterson, Schnell, \& Doyle, {\it Nature} {\bf 497}, 475 (2013) introduced a microwave experiment that produces a perpendicularly polarized molecular emission when chiral molecules are resonantly excited in the presence of a Stark Field that is then adiabatically switched off before observation of the emission.   The sign of the signal is opposite for two stereo-enantiomers and thus the magnitude of the signal gives the enantiomeric excess of the sample.    This paper presents a detailed presentation of the theory behind this and provides expressions for the absolute calculation of the expected signal strength for different transitions.  
\end{abstract}
 
 \maketitle

 Since Pasteur recognized in 1849 that optical rotation arose from what we now call chirality on a molecular scale~\cite{Flack09}, the topic of chirality has been central to chemistry~\cite{Mislow02}.  Pasteur also recognized that all chiral compounds uniquely arise from biological sources~\cite{Haldane60}, which likely played a role in motivating his later work experimentally disproving spontaneous generation.   The relationship of molecular chirality to the underlying symmetries of physical laws is discussed in Barron~\cite{Barron86} and Quack~\cite{Quack89, Quack01}. How chiral life arose from a largely achiral chemical soup remains one of the fundamental questions of modern science~\cite{Guijarro60}.   The continued importance of Chirality to modern Chemistry is demonstrated by a recent search for "enantiomer" in a scientific bibliographic database which returned nearly 10,000 publications in the last five years alone~\cite{WOS}.   Despite this, the typical practicing molecular spectroscopist almost never directly deals with molecular chirality in his or her work because the optical properties of enantiomers are almost identical.  The development of sensitive spectroscopic methods to distinguish enantiomers has long been a goal of many spectroscopists.  

In an exciting recent publication, Patterson, Schnell, \& Doyle~\cite{Patterson13} presented a microwave spectroscopy experiment that uniquely generates signals for chiral molecules and then only if there is an enantiomeric excess in the sample.  In their method, a chiral molecule is excited with linearly polarized microwave radiation resonant with an electric dipole transition while subjected to a Stark (DC Electric) Field oriented perpendicular to the polarization direction of the excitation field.   In chiral or achiral molecules, this excitation produces a free induction decay (FID) emission field that is polarized parallel to the excitation field.   However, uniquely for a chiral molecule, an FID is generated by the sample \textit{once the Stark field is switched off} that is polarized perpendicular to both the Stark and excitation field directions.   The signal is proportional, in the limit of weak excitation and Stark state mixing, to the product of the three projections of the dipole moment on the inertial axes of the molecule.   While one can define the orientation of the inertial axes to make any two the dipole moments positive, the sign of the third will be fixed if one demands a right-handed coordinate system.  This triple product of dipole moment projections has opposite sign for two enantiomers of a chiral molecule and is zero for any achiral molecule.  For any racemic sample, the perpendicularly polarized FID fields from the two enantiomers will destructively interfere, again producing no perpendicularly polarized signal. 

More recently,  Patterson \textit{et al.}\cite{Patterson13b, Patterson14}, Shubert \textit{at al}\cite{Shubert14a, Shubert14b}, and Lobsiger \textit{et al.}\cite{Lobsiger15}  have reported results for a related method where the DC Stark field is replaced by a resonant AC field, resulting in a chiral selective four wave mixing signal.  While there are obvious advantages of the four wave mixing approach, there are cases where the Stark mixing approach may be preferred.  For one, it is easier to use a microwave cavity to enhance the emitted power by the $Q$ of the resonator.  For another, it may prove easier to relate the absolute phase of the observed emission to the product of the three electric dipole moment components which can be used to determine which enantiomer is in excess in a sample, i.e. combined with predictions of Quantum chemistry, to determine the absolute configuration of a chiral sample.   Because the DC Stark mixing does not suffer from the same phase mismatching issues as the four wave mixing experiments, it may also proved easier to quantitatively relate the observed chiral signal strengths to the size of the enantiomeric excess ($ee$) of a sample, particularly for cases where does not have a sample of the same molecules with known $ee$ to use as a standard.  

 In this paper, I will present an analysis of Stark-Field induced Chiral signals.~\cite{Patterson13}  I will give results that will allow for the calculation of the size of the chiral emission, relative to the regular, achiral microwave free induction decay emission, for any molecule given its rotational constants and dipole moment projections.  We will start in the weak field limits where perturbation theory can be used.   This will be followed by analysis of the cases of both strong Stark mixing and also pulses that produce maximal coherence in the sample.     

\section{Signals in the perturbative limit}

Consider a rigid chiral, asymmetric top molecule with rotation constants $A$, $B$, and $C$ and projection of the molecular electric dipole along these axes of $\mu_{a}$, $\mu_{b}$ and $\mu_{c}$ respectively.  Most chiral molecules have no symmetry operations beyond the identity (point group $C_{1}$) and for these one expects all three components of $\mu$ to be nonzero, though obviously symmetry does not specify the relative sizes of the components.  Some of the simplest chiral molecules, such as $H_2O_2$, have a $C_2$ or higher axis of symmetry, which is necessarily along one of the principle axes and the permanent dipole moment is also along this symmetry axis.  As such, the Patterson \textit{et al.}\cite{Patterson13, Patterson13b} methods of chiral detection will not work for these molecules.

The zero field rotational levels of an asymmetric top\cite{Townes55} can be written, in increasing energy order at fixed total angular momentum quantum number $J$, with labels $|J,\tau, M>$ where $\tau = -J \ldots J$ and $M$ is the projection of $J$ on the laboratory $Z$ axis.  The more common $K_{p}$, $K_{o}$ labels, where $p$ is shorthand for prolate and $o$ oblate limits, can be calculated using $K_{p} =$ Floor$\left( \frac{\tau+J+1}{2}    \right)$
and $K_{o} =$ Floor$\left( \frac{J + 1 - \tau}{2}    \right)$.  The field-free rigid rotor hamiltonian is invariant to reflection in the planes normal to the inertial axes and has symmetry of the Klein 4-group (Vierergruppe).  The energy  eigenfunctions transform under the symmetry operations as one of four irreducible representations that are defined by whether $K_{p}$ and $K_{o}$ are even or odd integers.  The wavefunctions can be written as linear combinations of symmetric top functions:
\begin{equation}
\left|J,\tau, M \right>  = \sum_{K = -J}^{J} A_{K,\tau} \left|J, K, M \right>
\end{equation}
where $\left|J,K,M \right>$ are symmetric top functions  which are functions of the Euler angles that give the orientation of the inertial axes relative to the laboratory fixed axes and can be expressed explicitly in terms of Wigner $D$ matrices~\cite{Zare}.  $K$ is the quantum number for projection of angular momentum on the molecular fixed inertial unique axis (usually either $A$ or $C$) and $M$ is the projection of $J$ on the laboratory fixed $Z$ axis.
The values of $A_{K,\tau}$ only depend upon Ray's asymmetry parameter $\kappa = (2B - A - C)/(A-C)$.\cite{Townes55}   Likewise, the energy eigenvalues in zero field can be written $E(J,\tau) = \frac{1}{2}( A + C) J (J+1) + \frac{1}{2} (A - C) E_{J\tau}(\kappa)$, where $E_{J\tau}(\kappa)$ are the energy eigenvalues for the rigid rotor Hamiltonian with rotational constants $A = 1$, $B = \kappa$, and $C = -1$.   

Before considering the general case, it is instructive to first consider the particular case of driving the $0_{00} \rightarrow 1_{10}$ transition as was done experimentally by Patterson, Schnell, \& Doyle~\cite{Patterson13}.  We select our axis system such that the Stark Field is along the laboratory $Z$ axis, the excitation pulse is polarized along the $X$ direction and the chiral signal will be polarized along the $Y$ direction.  Note that this is different from what was used in \cite{Patterson13} but this change has no physical consequence and I find this choice simplifies the analysis somewhat as then $M$ remains a good quantum number.  We neglect $\Delta J = \pm 1$ mixing terms in the Stark effect.  With this approximation, there are no Stark contributions to the $0_{00}$ state.  The $1_{10}$ state has Stark Matrix elements with the $1_{11}$ state with an $a$ polarized dipole and energy spacing $B-C$ and state $1_{01}$ with a $b$ polarized dipole and spacing $A-C$.   We will use the notation $\left|J,\tau,M\right>'$ to denote the energy eigenstate in a Stark Field that adiabatically connects with the field free $\left|J, \tau, M \right>$ state.  To first order in the Stark Field we have:
\begin{equation}
\left| 1_{10,M} \right>' = \left| 1_{10,M} \right> - \frac{E_{\rm S} \mu_{a}}{B-C} \left| 1_{11,M} \right> \left< 1_{11,M} | 
\phi_{aZ} | 1_{10,M} \right>  - \frac{E_{\rm S} \mu_{b}}{A-C} \left| 1_{01,M} \right> \left< 1_{01,M} | 
\phi_{bZ} | 1_{10,M} \right> 
\end{equation}
where $\phi_{\alpha, G}$ are the direction cosine matrix elements between the molecular fixed axes $\alpha = a, b, c$ and the laboratory fixed axes $G = X,Y, Z$.   
The sample is driven by an excitation pulse, $E_{\rm ex} \cos ( \omega t)$, that is on resonance with the degenerate $\left|0_{00;0} \right>' \rightarrow  \left| 1_{10,\pm 1}\right>'$ transitions. 
For times after the end of a weak excitation pulse of length $\Delta t$, and including only near resonance contributions, the wavefunction is
\begin{eqnarray}
\left|\psi(t \right> &=& \left| 0_{00} \right>' + \sum_{M = \pm 1} a_{M} \left| 1_{10;M}\right>' \exp \left( - i E'( 1_{10}) t /\hbar  )     \right) \nonumber \\
a_{M = \pm 1} &=& \frac{i E_{\rm ex} \Delta t }{2\hbar} \left[   \mu_{c}  \left< 1_{10,M}| \phi_{cX}| 0_{00} \right>                      
-\frac{E_{\rm S} \mu_{a}\mu_{b}}{B-C}  \left< 1_{10,M}| \phi_{aZ}| 1_{11,M} \right>   \left< 1_{11,M}| \phi_{bX}| 0_{00} \right>    \right. \nonumber \\
& & \hspace{1in} \left. -\frac{E_{\rm S} \mu_{a}\mu_{b}}{A-C}  \left< 1_{10,M}| \phi_{bZ}| 1_{01,M} \right>   \left< 1_{01,M}| \phi_{aX}| 0_{00} \right>     \right] \nonumber \\
&=&  \frac{i E_{\rm ex} \Delta t }{2 \hbar} \left[   \mu_{c}  \left(  \pm \frac{i}{\sqrt{6}}   \right)     
-\frac{E_{\rm S} \mu_{a}\mu_{b}}{B-C}  \left(  \pm \frac{1}{2}   \right)   \left(  \pm \frac{1}{\sqrt{6}}    \right) 
-\frac{E_{\rm S} \mu_{a}\mu_{b}}{A-C}  \left(  \pm \frac{1}{2}   \right)   \left(  \mp \frac{1}{\sqrt{6}}    \right)    
\right]         \nonumber \\
&=&  \frac{i E_{\rm ex} \Delta t }{4 \sqrt{6} \hbar} \left[  \pm 2i \mu_{c} - \frac{E_{\rm S} \mu_{a}\mu_{b}(A-B)}{(A-C)(B-C)}    
\right]
\end{eqnarray}
Note that the contributions from mixing of the $1_{11}$ and $1_{01}$ states with the $1_{10}$ state are opposite in sign.
We assume that the Stark Field is turned off adiabatically before the FID of the sample dephases.
For detection with field off, we have polarization in the Y direction of
\begin{eqnarray}
P_{Y} &=& \Delta N \sum_{M = \pm 1} a(M) \exp \left( - i E( 1_{10}) t /\hbar \right)   \left< 0_{00} | \mu_{c} \phi_{cY} | 1_{10;M}\right>  + \text{c.c.} \nonumber \\
&=& \Delta N \sum_{M = \pm 1}  \frac{i E_{\rm ex} \Delta t }{4 \sqrt{6} \hbar} e^{-i (A+B)t/ \hbar}  \left[  \pm 2i \mu_{c} - \frac{E_{\rm S} \mu_{a}\mu_{b}(A-B)}{(A-C)(B-C)}    
\right] \left(   \mu_{c} \left( - \frac{1}{\sqrt{6}}    \right)    \right) + \text{c.c.} \nonumber \\
&=& i \Delta N \left(  \mu_{a} \mu_{b} \mu_{c} \right)  \frac{E_{\rm S} E_{\rm ex} \Delta t }{12 \hbar} \frac{A-B}{(A-C)(B-C)}
e^{-i (A+B)t/\hbar} + \text{c.c.}
\end{eqnarray}
where $\Delta N = N(0_{00}) - N(1_{10})/3$ is the thermal difference in number density for the transition and c.c. stands for complex conjugate of what comes before.  As predicted, the polarization to zero order in $E_{\rm S}$ is zero because of the cancelation in the sum over $M = \pm 1$.   The Stark induced signal is proportional to the product of $ \mu_{a} \mu_{b} \mu_{c}$ and thus is of opposite sign for two enantiomers.  Thus, if we have a mixture of enantiomers, the net amplitude of the polarization will be proportional to the difference in the number density of the $R$ and $S$ forms, i.e. proportional to the enantiomeric excess.
We can compare this signal to the $X$ polarized polarization that leads to the standard FID:
\begin{eqnarray}
P_{X} &=& \Delta N \sum_{M = \pm 1} a(M) \exp \left( - i E( 1_{10}) t /\hbar \right)   \left< 0_{00} | \mu_{c} \phi_{cX} | 1_{10;M}\right>  + \text{c.c.} \nonumber \\
&=&  \Delta N \sum_{M = \pm 1}  \frac{i E_{\rm ex} \Delta t }{4 \sqrt{6} \hbar} e^{-i (A+B)t/ \hbar}  \left[  \pm 2i \mu_{c} - \frac{E_{\rm S} \mu_{a}\mu_{b}(A-B)}{(A-C)(B-C)}    
\right] \left(   \mu_{c} \left( \mp \frac{i}{\sqrt{6}}    \right)    \right) + \text{c.c.} \nonumber \\
&=&  i \Delta N  \mu_{c}^2  \frac{E_{\rm ex} \Delta t }{6 \hbar} 
e^{-i (A+B)t/\hbar} + \text{c.c.}
\end{eqnarray}
Both $P_Y$ and $P_X$ are phase shifted by $90$ degrees from the excitation wave.  The FID field emitted these polarizations will be  further shifted by $90$ degrees, so, both net fields will be out of phase with the driving field, should it had been continued into the detection interval.  This result can be physically understood if one considers an experiment that seeks to simultaneously detect both $P_X$ and $P_Y$.   Imagine that after the sample (in the direction of propagation of the excitation wave, which is $Z$) we place a horn capture the $X$ polarized FID, and then we rotate it to detect the $Y$ polarized wave.   The sign of this wave (relative) to the $X$ polarized wave will change depending upon if we do a clockwise rotation  (looking in the direct of the wave propagation) or counter-clockwise rotation.  In the clockwise rotation case, we are defining $Y$ to form a right-handed coordinate system with respect to the previously defined $X$ and $Z$ axes, while the opposite rotation defines a left-handed coordinate system.   The net FID has rotated like in normal optical rotation.

Before we turn off $E_{\rm S}$, we have to calculate the transition of $\left| 0_{00} \right>$ to the Stark Field mixed state.   That leads to additional contributions to $P_{Y}$:
\begin{eqnarray}
P_{Y} &=& \Delta N \sum_{M = \pm 1} a(M) e^{-i (A+B)t} \left[ \mu_{c} \left< 0_{00} |  \phi_{cY} | 1_{10;M}\right> 
-\frac{E_{\rm S} \mu_{a}\mu_{b}}{B-C}  \left< 0_{00}| \phi_{bY}| 1_{11,M} \right>   \left< 1_{11,M}| \phi_{aZ}| 1_{10;M} \right>    \right. 
 \nonumber \\
 & & \hspace{1in} \left. -\frac{E_{\rm S} \mu_{a}\mu_{b}}{A-C}  \left< 0_{00}| \phi_{aY}| 1_{01,M} \right>   \left< 1_{01,M}| \phi_{bZ}| 1_{10;M} \right>     \right] + \text{c.c.} \nonumber \\
&=& \Delta N \sum_{M = \pm 1} a(M) e^{-i (A+B)t} \left[  \mu_{c} \left( - \frac{1}{\sqrt{6}}    \right)    
- \frac{E_{\rm S}\mu_{a}\mu_{b}}{B-C} \left( - \frac{i}{\sqrt{6}}    \right)   \left( \pm \frac{1}{2}    \right)   \right. \nonumber \\
&& \hspace{1in} \left. - \frac{E_{\rm S}\mu_{a}\mu_{b}}{A-C} \left(    \frac{i}{\sqrt{6}}    \right)   \left( \pm \frac{1}{2}    \right) 
\right] + \text{c.c.} \nonumber \\
\end{eqnarray}
substituting in from above for $a(M)$ gives 
\begin{eqnarray}
P_{Y} &=&\frac{ i \Delta N E_{\rm ex} \Delta t}{12 \hbar}  e^{-i (A+B)t}\sum_{M = \pm 1}
(\mp 1) \left[  \mu_{c} \mp i \frac{E_{\rm S} \mu_{a} \mu_{b}  (A-B) }{2 (A-C)(B-C) }  \right] \left[  \mu_{c} \pm i \frac{E_{\rm S} \mu_{a} \mu_{b}  (A-B) }{2 (A-C)(B-C) }  \right]  \nonumber \\
& & \hspace{1in} +  \text{c.c.} \nonumber \\
 &=&\frac{i  \Delta N E_{\rm ex} \Delta t}{12 \hbar} e^{-i (A+B)t} \sum_{M = \pm 1} (\mp 1) \left[ \mu_{c}^{2} +    \left(\frac{E_{\rm S} \mu_{a} \mu_{b}  (A-B) }{2 (A-C)(B-C) }   \right)^{2}  \right] + \text{c.c.} = 0
\end{eqnarray}
and thus demonstrating that the $Y$ polarized signal (i.e. perpendicular to the excitation polarization) does not appear until the Stark field is changed.  
Note that if we had inverted the direction of the Stark field, which changes the sign $E_{\rm S}$, we would double the amplitude of the chiral FID compared to simply shorting the field to zero.

Before considering the effect of excitation on a generic rotational transition in the presence of a Stark Field, let us consider the field free excitation of a transition $\left|J'',\tau''\right> \rightarrow \left|J',\tau'\right>$ with $X$ polarized radiation.  We assume that we start with a thermal population  with states $\left|J,\tau, M\right>$ equally populated for fixed $J, \tau$, with total population in a level $J, \tau$ of $N(J, \tau)$.  The excitation pulse will be assumed to have electric field amplitude $E_{\rm ex}$, duration $\Delta t$ and an angular frequency detuning from exact resonance $\Delta \omega$.    $\alpha = a, b$, or $c$ are the directions of the transition moments in the principle axis system of the rigid molecule.   Following excitation, we will generate a sample polarization in the $Y$ directions 
\begin{eqnarray}
P_{Y} &=& \Delta N \left(  \frac{E_{\rm ex} \Delta t \left(  1 - e^{-i \Delta \omega \Delta t}     \right)}{2 \hbar \Delta \omega} \right) e^{-i \omega_{0} t /\hbar} \times \nonumber \\
&  & \sum_{\alpha, M'', M' = M'' \pm 1} \mu_{\alpha}^{2}
\left< J'',\tau'',M''| \phi_{\alpha, Y} | J', \tau', M' \right> \left<  J', \tau', M' |  \phi_{\alpha, X} |  J'',\tau'',M''     \right> + \text{c.c.}
\end{eqnarray}
where $\omega_0 = ( E(J',\tau') - E(J'',\tau'') )/\hbar$ is the angular frequency of the emission and
\begin{equation}
\Delta N = \left( \frac{N(J'',\tau'')}{2J''+1} - \frac{N(J',\tau')}{2J'+1} \right)
\end{equation}
is again the thermal population difference between states of fixed $M$.
We can decompose the direction cosine matrix elements as:
\begin{eqnarray}
\left< J, \tau, M | \phi_{\alpha, G} | J', \tau', M'\right>  =  \left<J,\tau | \phi_{\alpha} | J', \tau' \right> \left<J,M | \phi_G |J',M'\right> \\
\left< J, \tau  | \phi_{\alpha} | J', \tau' \right>  =  \left<J|\phi_J|J'\right> \sum_{ K,K' } A_{K,\tau} A_{K',\tau'} \left<J,K|\phi_{\alpha}|J',K'\right>
\end{eqnarray}
where the matrix elements of $\phi_J, \phi_{J,K}$, and $\phi_{J,M}$ are defined in Table 4.2 of Townes \& Schawlow~\cite{Townes55}.  For a given transition, at most one value of $\alpha$ will give nonzero matrix elements $\left< J'',\tau'',M''| \phi_{\alpha, G} | J', \tau', M' \right>$.      We can use the symmetries
\begin{eqnarray}
\left<  J', \tau', M''  \pm 1|  \phi_{\alpha, Y} |  J'',\tau'',M''     \right> &=& \pm i  \left<  J', \tau', M'' \pm 1 |  \phi_{\alpha, X} |  J'',\tau'',M''     \right> \\
\left<  J', \tau', -M''  \mp 1|  \phi_{\alpha, X} |  J'',\tau'', -M''     \right> &=& (-1)^{|J' - J''|}  \left<  J', \tau', M'' \pm 1 |  \phi_{\alpha, X} |  J'',\tau'',M''     \right> 
\end{eqnarray}
to show that
\begin{eqnarray}
&&\left< J'',\tau'',M''| \phi_{\alpha, Y} | J', \tau', M'' \pm 1 \right>\left<  J', \tau', M'' \pm 1 |  \phi_{\alpha, X} |  J'',\tau'',M''     \right>
=  \nonumber \\
&&\hspace{0.2in} - \left< J'',\tau'',-M'' | \phi_{\alpha, Y} | J', \tau', -M'' \mp 1 \right>\left<  J', \tau', -M'' \mp 1 |  \phi_{\alpha, X} |  J'',\tau'', -M''     \right> \label{sym.eq}
\end{eqnarray}
Thus, as long as the population of the states with $M'' =+M$ and $M'' = -M$ are equal, the terms with $M'' = M, M' = M \pm 1$ will cancel those with $M'' = -M, M' = -M \mp 1$ and we get no net polarization perpendicular to the excitation polarization and thus no  $Y$ polarized contribution to the Free Induction Decay (FID) emission.  This result is independent of the labels we give to the axes of excitation and detection.  Physically, this is a consequence of the fact that  the sample is not birefringent if we have zero orientation of the sample.  Even achiral molecules in states of definite $M$ are optically active, i.e. have different susceptibilities for left and right handed light.  Such individual states have what Barron calls ``false chirality''~\cite{Barron86} but not chirality, based upon the symmetry properties under time reversal.

Consider application of a DC Stark field along the $Z$ axis with electric field $E_{\rm S}$, and for now assume it is small enough that we can use perturbation theory.  To first order, we get for the wavefunctions in the presence of the field:
\begin{eqnarray}
|J,\tau,M>' &=&  |J, \tau, M> \\
&&  - \sum_{J_{s} = J - 1}^{J+1} \sum_{\tau_{s}} \sum_{\alpha} \frac{\mu_{\alpha} E_{\rm S}}{(E(J,\tau) -E(J_{s},\tau_{s}))} 
\left| J_{s},\tau_{s},M \right>\left<  J_{s}, \tau_{s}, M | \phi_{\alpha,Z} | J, \tau, M \right> \nonumber
 \end{eqnarray}   

We will now consider excitation on a transition $\left| J'', \tau'', M'' \right>' \rightarrow \left|J', \tau', M' \right>'$ with, as before, a resonant field with amplitude $E_{\rm ex}$ polarized along the $X$ axis that is nonzero for time interval $\Delta t$ and that the excitation is sufficiently weak we can use first order time dependent perturbation theory.   Following the excitation pulse, the wavefunction for each initially Stark perturbed rotational state $\left| J'', \tau'', M'' \right>'$
can be written:
\begin{equation}
\left| \psi(t, M'') \right> = \left| J'', \tau'', M'' \right>' e^{ -i E(J'',\tau'', M'')/\hbar} + \sum_{M'} a(M'',M') \left| J', \tau', M' \right>' e^{ -i E(J',\tau', M')/\hbar }
 \end{equation}
 Making the rotating wave approximation, we can write for times after the end of the excitation pulse that:
\begin{equation}
a(M'',M') =   - E_{\rm ex} \frac{  e^{-i \Delta\omega(M',M'')  \Delta t} - 1  }{  2 \hbar ( \Delta \omega(M',M'') )}
\sum_{\beta}  \left<  J', \tau', M'  \right|' \mu_{\beta} \, \phi_{\beta,X} \left|  J'', \tau'', M''  \right>' 
\end{equation}
where $\Delta \omega(M',M'')$ is the $M$ dependent detuning for the transition.  Due to the perpendicular polarization, we have nonzero contributions only for $M' = M'' \pm 1$.  To simplify things, we assume that the second order Stark splitting of the transition is sufficiently small that $\Delta \omega(M',M'') \Delta t << 1$, in which case we can treat all $M$ components as on resonance.   This allows us to write
\begin{equation}
a(M'',M') = i \frac{ E_{\rm ex} \Delta t}{2 \hbar} \sum_{\beta}   \left<    J', \tau', M'  \right|' \mu_{\beta} \, \phi_{\beta,X} \left|  J'', \tau'', M''  \right>' 
\end{equation}
More generally, we can replace $E_{\rm ex} \Delta t$ by $\int E_{\rm ex}(t)dt$.
We will keep only terms up to linear in $E_{\rm S}$.  This allows us to write for the transition matrix elements
\begin{eqnarray}
& & \sum_{\beta} \mu_{\beta}\left<  J', \tau', M'  \right|' \phi_{\beta,X} \left|  J'', \tau'', M''  \right>'  =
\sum_{\beta} \mu_{\beta} \left<  J', \tau', M'  \right| \phi_{\beta,X} \left|  J'', \tau'', M''  \right> \nonumber \\
  && \hspace{1in} -  E_{\rm S} 
  \sum_{\alpha, \beta, J_{s}, \tau_{s}}   \mu_{\alpha} \mu_{\beta} \left( \frac{ \left<  J', \tau', M' \right| \phi_{\alpha,Z} \left| J_{s}, \tau_{s}, M' \right>    
  \left<  J_{s}, \tau_{s}, M' \right| \phi_{\beta,X} \left|  J'', \tau'', M''  \right>   }{E(J' ,\tau') - E(J_{s},\tau_{s})} \right. \nonumber \\ 
  && \hspace{1.5in} + \left.
  \frac{ \left<  J', \tau', M' \right| \phi_{\beta,X} \left| J_{s}, \tau_{s}, M'' \right>    
  \left<  J_{s}, \tau_{s}, M'' \right| \phi_{\alpha,Z} \left|  J'', \tau'', M''  \right>   }{E(J'', \tau'') - E(J_{s},\tau_{s})} \right)
\end{eqnarray}
Since we adiabatically turn off the Stark Field before detecting the emission polarized along $Y$, we calculate the emission with $\left|J,\tau,M\right>'$
replaced by $\left|J,\tau,M\right>$, {\it i.e.} the zero order wavefunctions.  This leads to a molecular polarization polarized along $Y$, $P_{Y}$, that results in emitted FID field oscillating at the excitation frequency and polarized along $Y$,
\begin{equation}
P_{Y} = \Delta N \sum_{M',M'',\gamma} a(M'',M') e^{-i \omega_0 t}  \mu_{\gamma} \left< J'', \tau'', M'' \right| \phi_{\gamma,Y} \left|J', \tau', M'\right> + \text{c.c.}
\end{equation}

Above, it was demonstrated that the terms zero order in $E_{\rm S}$ give zero contribution after the sum over $M$ levels.
The terms linear in $E_{\rm S}$ give:
\begin{eqnarray}
&&P_{Y}  = \frac{ - i E_{\rm S} E_{\rm ex}\Delta t \Delta N}{2 \hbar } e^{-i \omega_0 t} \sum_{\alpha, \beta, \gamma} \mu_{\alpha} \mu_{\beta} \mu_{\gamma}
 \sum_{M',M'',J_{s},\tau_{s}} \left<J'',\tau'' | \phi_{\gamma} | J',\tau'\right> \left<J'',M''|\phi_{Y}|J',M'\right> \times \nonumber \\
 & &    \left( \frac{  \left<J',\tau'|\phi_{\alpha}|J_{s},\tau_{s}\right> \left<J',M'|\phi_{Z}|J_{s},M'\right>    
 \left<J_{s},\tau_{s}| \phi_{\beta}|J'',\tau''\right> \left<J_{s},M'|\phi_{X}|J''M''\right> }{E(J' \tau') - E(J_{s},\tau_{s})} +
 \right. \nonumber \\
 & & \left.      \frac{  \left<J'\tau'|\phi_{\beta}|J_{s},\tau_{s}\right> \left<J'M'|\phi_{X}|J_{s},M''\right>    
 \left<J_{s}, \tau_{s} | \phi_{\alpha}|J''\tau''\right> \left<J_{s}, M''|\phi_{Z}|J'', M''\right> }{E(J'' ,\tau'') - E(J_{s},\tau_{s}) }
 \right) + \text{c.c.}
\end{eqnarray}
Given the selection rules for an asymmetric top, we can have a cycle of transitions $(J'',\tau'') \rightarrow (J_{s},\tau_{s}) \rightarrow (J' ,\tau') \rightarrow (J'',\tau'')$ only if $\alpha \ne \beta \ne \gamma$, in which case $\alpha, \beta, \gamma$ are a permutation of $a, b, c$. The $P_Y$ signal will only be nonzero if all $\mu_{a}, \mu_{b}, \mu_{c} \ne 0$, which is true only for molecules of $C_1$ symmetry, which are chiral.  Only $M' = M'' \pm 1$ give nonzero contributions. 

We can do the sums over $M$ quantum numbers if we specify the $J$ quantum numbers.
For an R branch transition $J'' = J,  J'  = J + 1$, $J_{s} = J, J+1$.  we get after summing over $M$ quantum numbers:
\begin{eqnarray}
&&P_{Y} \left( J ,\tau''  \rightarrow J+1, \tau'  \right)  = \nonumber \\
&&  \frac{ 2 E_{\rm S} E_{\rm ex}\Delta t  \Delta N \left(   \mu_{a} \mu_{b} \mu_{c} \right)}{3 \hbar }e^{-i \omega_0 t} (J+1)(2J+1)(2J+3)
\sum_{\alpha, \beta, \gamma, \tau_{s}}   \left<J \tau''|\phi_{\gamma}|J+1, \tau'\right> 
 \nonumber \\
 & &     
 \times  \left[  - \left<J+1, \tau' | \phi_{\alpha} | J+1,\tau_{s}\right>\left<J+1, \tau_{s}| \phi_{\beta} | J , \tau'' \right>
 \left( \frac{J+2}{ E_{J+1, \tau'} - E_{J+1, \tau_{s}}} -  \frac{J+2}{ E_{J, \tau''} - E_{J+1, \tau_{s}}} \right) \right. \nonumber \\
&  & \left.
+ \left<J+1, \tau' | \phi_{\alpha} | J,\tau_{s}\right>\left<J, \tau_{s}| \phi_{\beta} | J , \tau'' \right>
\left( \frac{J}{ E_{J+1, \tau'} - E_{J, \tau_{s}}} -  \frac{J}{ E_{J, \tau''} - E_{J, \tau_{s}}} \right) \right] + \text{c.c.}
\end{eqnarray}

For a Q branch transition with $J' = J'' =J$, $J_{s} = J-1, J, J+1$.  Summing over the $M$ quantum numbers gives the following result
\begin{eqnarray}
&&P_{Y} \left( J ,\tau''  \rightarrow J, \tau'  \right)  = \nonumber \\
&&  \frac{ 2 E_{\rm S} E_{\rm ex}\Delta t  \Delta N \left(   \mu_{a} \mu_{b} \mu_{c} \right)}{3 \hbar }e^{-i \omega_0 t} J(J+1)(2J+1)
\sum_{\alpha, \beta, \gamma, \tau_{s}}   \left<J \tau''|\phi_{\gamma}|J, \tau'\right> 
 \nonumber \\
 & &     
 \times  \left[  \left<J, \tau' | \phi_{\alpha} | J+1,\tau_{s}\right>\left<J+1, \tau_{s}| \phi_{\beta} | J , \tau'' \right>
 \left( \frac{2J+3}{ E_{J, \tau'} - E_{J+1, \tau_{s}}} -  \frac{2J+3}{ E_{J, \tau''} - E_{J+1, \tau_{s}}} \right) \right. \nonumber \\
&  & 
- \left<J, \tau' | \phi_{\alpha} | J,\tau_{s}\right>\left<J, \tau_{s}| \phi_{\beta} | J , \tau'' \right>
 \left( \frac{1}{ E_{J, \tau'} - E_{J, \tau_{s}}} -  \frac{1}{ E_{J, \tau''} - E_{J, \tau_{s}}} \right) \nonumber  \\
& & \left.
- \left<J, \tau' | \phi_{\alpha} | J,-1\tau_{s}\right>\left<J-1, \tau_{s}| \phi_{\beta} | J , \tau'' \right>
 \left( \frac{2J-1}{ E_{J, \tau'} - E_{J-1, \tau_{s}}} -  \frac{2J-1}{ E_{J, \tau''} - E_{J-1, \tau_{s}}} \right) 
 \right] + \text{c.c.}
\end{eqnarray}

The triple product of direction cosine matrix elements will be imaginary, which means that the FID in the $Y$
polarization will be phase shifted by $\pm \pi$ from the driving radiation, like the $X$ polarized polarization, so both
will lead an emitted field out of phase with the driving radiation. 
Each of the direction cosine elements $\left< \phi_{\alpha} \right>$ scales $\sim J^{-1}$, so
except for the $J'' = J' = J_s$ terms, we get a net $\sim J$ scaling of the terms, which reflects
the degeneracy of states. In most cases, the dominant terms will be Stark Mixing of asymmetry doubles
as these have the smallest energy differences.  For such mixing, the chiral signal on Q branch transitions will be weak relative
to R branch transitions.

It is worth noting that the sum over $M$ quantum numbers in the triple product of direction cosine factors gives zero unless the three laboratory projects are a permutation of $X,Y,Z$,  i.e. that the polarizations of the Stark, Excitations, and Detection fields are each perpendicular.   This reflects the fact that the chiral signal arises from projections of the three dipole moment components from the rotating molecular frame onto a laboratory frame defined by the three electric field directions.  If we project the three vectors onto a plane, then there is no way to determine if we are projecting from a left-handed or right-handed axis system formed by the three dipole projections on the molecular axis system.   

While the Stark Field is unchanged from its value during the excitation, the $Y$ polarization is calculated using the Stark perturbed wavefunctions:
\begin{equation}
P_{Y} = \Delta N  \sum_{M',M'',\gamma} a(M'',M') e^{-i \omega_0 t}  \mu_{\gamma} \left< J'', \tau'', M'' \right|' \phi_{\gamma,Y} \left|J', \tau', M'\right>' + \text{c.c.}
\end{equation}
This leads to additional terms which are almost the same as above but when summing over $M$ quantum numbers we get terms
\begin{eqnarray}
&&\sum_{M'',M' = M''\pm 1} \left< J'', M''| \phi_{Y}|J',M'' \pm 1 \right> \left< J', M'' \pm 1 | \phi_{X}|J_{s},M''\right> 
 \left< J_{s}, M'' \pm 1 | \phi_{Z}|J'',M''\right> + \nonumber \\
&& \left< J'', M''| \phi_{X}|J',M'' \pm 1 \right> \left< J', M'' \pm 1 | \phi_{Y}|J_{s},M''\right> 
 \left< J_{s}, M'' \pm 1 | \phi_{Z}|J'',M''\right>
\end{eqnarray}
and 
\begin{eqnarray}
&&\sum_{M'',M' = M''\pm 1} \left< J'', M''| \phi_{Y}|J_{s},M'' \pm 1 \right> \left< J_{s}, M'' \pm 1 | \phi_{Z}|J',M'' \pm 1 \right> 
 \left< J', M'' \pm 1 | \phi_{X}|J'',M''\right> + \nonumber \\
 && \left< J'', M''| \phi_{X}|J_{s},M'' \pm 1 \right> \left< J_{s}, M'' \pm 1 | \phi_{Z}|J',M'' \pm 1 \right> 
 \left< J', M'' \pm 1 | \phi_{Y}|J'',M''\right>
 \end{eqnarray}
where in both cases, only the first term in each sum is present when we detect with the Stark field off.  Using Eq.\,\ref{sym.eq}, it is evident that the sum of the first and second  terms in the above equations are identically zero, i.e. no signal exists until the Stark Field is changed and the amplitude of the FID will be proportional to the change in $E_{\rm S}$.  We could also do the excitation with $E_{\rm S} = 0$ and the detection with finite $E_{\rm S}$.  This will produce a signal of the same magnitude but changed in phase.   The signal could be doubled in size by changing the sign of $E_{\rm S}$ between excitation and detection.
However, in these cases, the FID will contain dephasing from the different $M$ components of the transition.   In many cases, sufficient microwave power is available so that the excitation time $\Delta t$ will be significantly shorter than the dephasing time of the field free FID, and in these cases it is better to have the Stark dephasing with the shorter excitation time instead of the longer detection time. 

If we restrict the Stark mixing to only $\Delta J = 0$ coupling terms, the signal (up to scale factors) depends only on the asymmetry factor $\kappa$.   Numerical calculations demonstrate that this chiral  signal is nonzero for all $R$ branch transitions of the chiral molecule for typical values of $\kappa$.   As could be expected, the predicted signal is particularly larger for transitions to levels with small asymmetry splittings as then the Stark mixing is larger due to the small energy denominator.   In these cases, the pairs of transitions to the two states that make up the asymmetry doublet have nearly equal and opposite sign signals.  

Because the enantiomer identification depends upon the phase of the $Y$ polarized FID signal, one should account for the fact that the Stark shifts will lead to some dephasing and phase shifts of the FID that depend upon the time delay and period for $E_{\rm S}$ to be ramped down.   One also needs for the time dependence of $E_{\rm S}$ to be sufficiently slow for the states to evolve adiabatically.   This likely puts some limit on the magnitude of the Stark mixing.  If the Stark interaction is such that one gets substantial state mixing, then the Stark shifts are comparable to the energy levels splittings and the two conditions of small dephasing of the $M$ components and that one has adiabatic evolution of the rotational states will be incompatible.     Table I contains the predicted chiral signals for the $R$ branch transitions calculated for the case that $\kappa = -0.706$ which is the value appropriate for 1,2-propanediol~\cite{Lovas09}, the molecule used by Patterson, Schnell \& Doyle to experimentally demonstrate this effect~\cite{Patterson13}.  The column labeled $P(Y)$ should be multiplied by $\left( \frac{N(J'',\tau'')}{2J''+1} - \frac{N(J',\tau')}{2J'+1} \right) \left( \frac{ 8 \pi E_{\rm S} E_{\rm ex}\Delta t  \left(   \mu_{a} \mu_{b} \mu_{c} \right)}{3 (A-C) h^{2}c } \right)$ to put the numbers in absolute units.  The column labeled $S$ has the line strength factor for each transition, $ S = \sum_{M'', M', \alpha} | \left< J', \tau', M' | \phi_{\alpha,X}| J'', \tau'', M'' \right> |^{2}$ and is one of the factors contributing the strength of the $X$ polarized FID emitted by the sample.

\section{High Stark Field Mixing}
The energy spectrum of an asymmetric top consists mostly of nearly degenerate levels known as asymmetry doublets.   These correspond semiclassically to symmetric and antisymmetric combination of states rotating in opposite directions around either the $a$ or $c$ axes and the splittings of asymmetry doublets can be well approximated by semiclassical tunneling integrals between the semiclassical trajectories corresponding to classical rotation around these axes~\cite{Harter84}.   For such states, the Stark effect at low field is dominated by mixing of these asymmetry doublet states if the corresponding dipole component is nonzero, which is the case of the chiral molecules we are considering in this work.    We will now assume we have an $a$ type asymmetry doublet and we will select the direction of the $a$ axis to make $\mu_{a} > 0$ which we are free to do.  The results apply to $c$ type asymmetry doublets if we invert the $a$ and $c$ labels.  

We consider the transitions $\left| J, \pm K\right> \rightarrow\left| J+1, \pm (K+1)\right>$.  We will use symmetric top functions.   We write the asymmetry splitting in the $\left| J, \pm K\right>$ state as $2\Delta \nu''$ and that in the $\left| J+1, \pm (K+1)\right>$ as $2\Delta \nu'$.  Let $\nu_{0} = (B+C)J + (2A - B - C)(K +1/2)$ be the transition frequency without asymmetry splittings.  In zero Stark field, we have four transitions at  $\nu_{0} - \Delta E''/h - \Delta E'/h$ ($J_{K,J-K} \rightarrow J+1_{K+1, J-K+1}, b$ type), $\nu_{0} - \Delta E''/h + \Delta E'/h$ ($J_{K,J-K} \rightarrow J+1_{K+1, J-K}, c$ type), $\nu_{0} + \Delta E''/h - \Delta E'/h$ ($J_{K,J-K+1} \rightarrow J+1_{K+1, J-K+1}, c$ type), and  $\nu_{0} + \Delta E''/h + \Delta E'/h$ ($J_{K,J-K+1} \rightarrow J+1_{K+1, J-K}, b$ type).  

As we turn up a Stark Field, each of the two asymmetry doublets will split from each other.  Initially, we will get second order energy shifts of  $\pm [\mu_{a} K M E_{\rm S} / J(J+1)]^{2} (2 \Delta E)^{-1}$, with $+$ for the upper asymmetry component and $-$ for the lower.  As the Stark shifts grows relative to the zero field asymmetry splittings, the shifts in each doublet evolve to first order in $E_{\rm S}$ with energy values $h\nu_{JKM} = h\nu_{JK}  \mp \mu_a K M E_{\rm S} /  J(J+1)$, where $h\nu_{JK}$ is the mean energy of the $J,K$ asymmetry doublet.   The transitions that adiabatically evolve from the $b$ type transitions will loose intensity, going to zero intensity at sufficiently high $E_{\rm S}$. The transitions that evolve from the $c$ type transitions in zero field will strengthen, taking up the decreasing strength of the $b$ types.   In high Stark Field (i.e. when $|\mu_{a} E_{\rm S} KM | /J(J+1) >> \Delta E', \Delta E''$), we can ignore the asymmetry slitting and the eigenstates will be nearly pure symmetric top states as the difference in Stark energy will quench the tunneling.   Consider excitation on the Stark component $\left| J, K, M \right> \rightarrow\left| J+1, K+1, M+1\right> \,\, (K,M \ge 0)$.  This will be exactly degenerate with the transition  $\left| J, -K, -M \right> \rightarrow\left| J+1, -K-1, -M-1\right>$.  For high $|M|$, these are the strongest transitions in the Stark multiplet of the $J, K$ transition.  If we polarize these transition with $X$ polarized radiation, we will create excited state amplitudes proportional to:
\begin{eqnarray}
a(\pm K, \pm M) &=& \frac{i E_{\rm ex} \Delta t}{2 \hbar} \left(  <J+1, K+1, M+1 | \mu_b \phi_{b,X} + \mu_c \phi_{c,X}  | J,K,M >        \right) \label{eqHFa}  \\
&=&  \frac{i E_{\rm ex} \Delta t}{2 \hbar} b_{JK} b_{JM} \left[ \mu_{b} \mp i \mu_{c}      \right]  \nonumber \\
b_{JK} &=& \sqrt{ \frac{(J+K+1)(J+K+2)}{2(J+1)\sqrt{ (2J+1)(2J+3)   }     } }
\end{eqnarray}

The states that $\left| J,\pm K, \pm M\right>$ and $\left|J+1, \pm (K+1), \pm (M+1)\right>$ will adiabatically evolve to, as the Stark Field is reduced to zero amplitude, depends upon the sign of $KM$.   We will assume here that $KM > 0$ in both initial and final quantum states (which are the stronger transitions), which implies that the states will evolve to the lower asymmetry states, which are $\frac{1}{\sqrt{2}}\left( \left|J, K, \pm M\right> - \left|J -K, \pm M \right> \right)$ for the lower state and $\frac{1}{\sqrt{2}}\left( \left|J+1, K+1, \pm (M+1) \right> - \left|J+1, -K-1, \pm (M+1) \right> \right)$ for the upper state.  The linear combination of the two lower states will give for $X$ and $Y$ polarized emission
\begin{eqnarray}
P_X &=& \sum_{\pm} - \mu_{c} (i b_{JK} \mu_c ) ( \mp b_{JM} ) a(\pm K, \pm M) e^{-i \omega_0 t } + c.c. \nonumber \\
& = & \mu_c^2 \left( \frac{i E_{\rm ex} \Delta t}{ \hbar}\right) b_{JK}^2 b_{JM}^2 e^{-i \omega_0 t } + c.c. \\
P_Y &=& \sum_{\pm} - \mu_{c} (i b_{JK} \mu_c ) ( i b_{JM} ) a(\pm K, \pm M) e^{-i \omega_0 t } + c.c. \nonumber \\
&=&  \mu_{b} \mu_{c} \left( \frac{ i E_{\rm ex}\Delta t}{\hbar} \right) b_{JK}^2 b_{JM}^2 e^{-i \omega_0 t } + c.c
\end{eqnarray}
The strongest transitions for perpendicular excitation are when $J = K = M$ for which the factor $b_{JK}^2 b_{JM}^2 $ becomes
\begin{equation}
b_{JJ}^4 = \frac{2J+1 }{ 4(2J+3)  }
\end{equation}

We thus see that the amplitude of the $Y$ polarized emission saturates at high Stark field such that the ratio of  the amplitudes of $Y$ polarized emission to $X$ polarize emission is proportional to $\mu_b / \mu_c$.
If we had taken $K M < 0$, the pumped states will adiabatically evolve to the upper levels of the asymmetry doublet as the Stark Field is turned off.  This will give a zero field $Y$ polarized molecular emission with the same expression as above except for a sign change.  
Note that since we have selected the coordinate system to make $\mu_{a} > 0$, this phase of the $P_Y$ emission will remain opposite in sign for a pair of enantiomers and thus its size will proportional to the \textit{e.e.} of the sample.  

 There will likely be a substantial phase shift during the time that $E_{\rm S}$ is turned off, but this should effect $P_{X}$ equally as $P_{Y}$ and thus the relative phase of the these signals will give the chirality.   Figure 1 shows a plot of the relative excitation amplitudes for the $b$ and $c$ type transitions of  $J'' = K'' = M'' = 2 \rightarrow J' = K' = M' =3 $ transitions of propanediol, which has a center transition frequency of 46 GHz as a function of $E_{\rm S}$, using the constants given by Lovas \textit{et. al.}~\cite{Lovas09}.

\begin{figure}[htbp]
\begin{center}
\includegraphics[width=17cm]{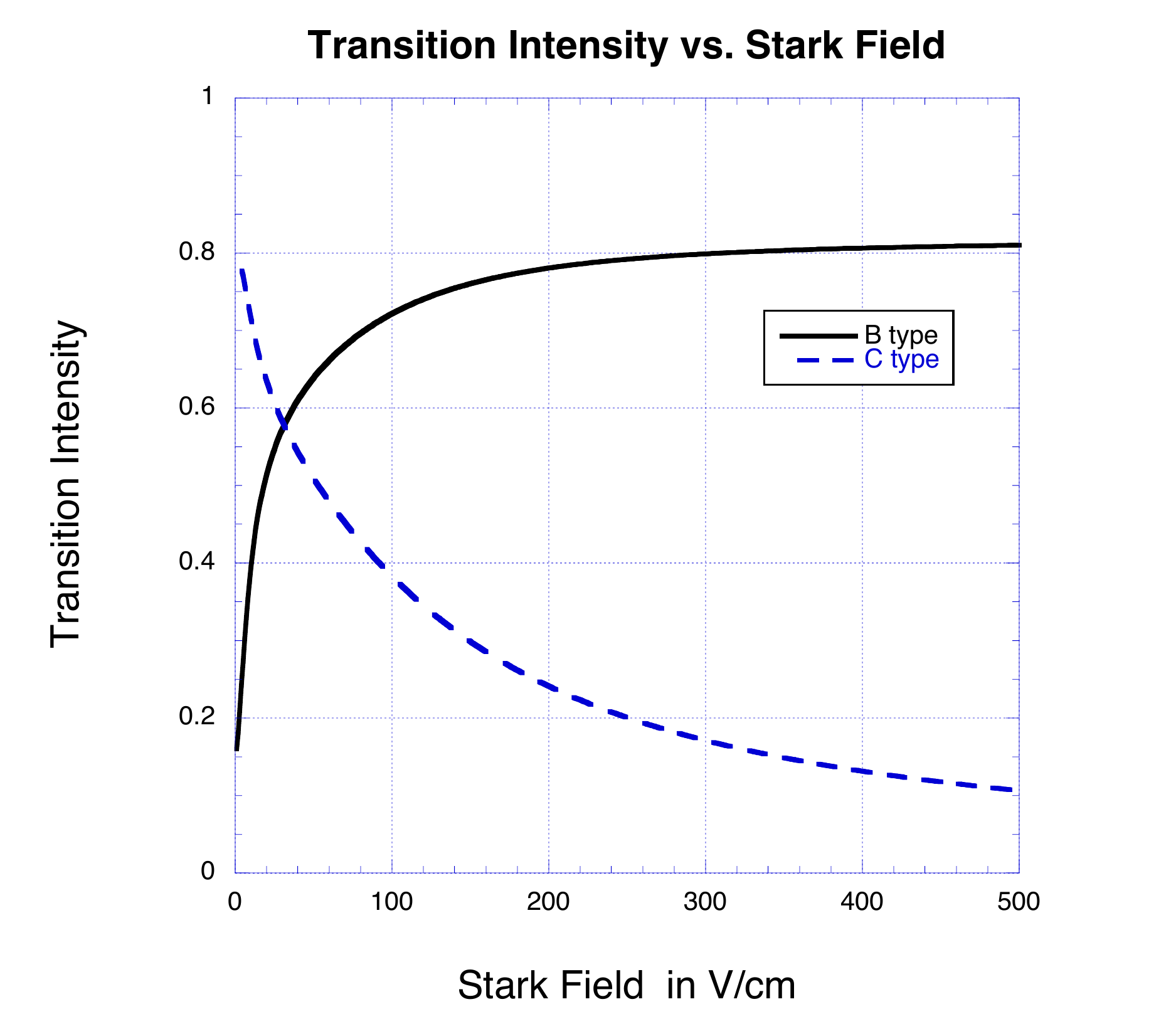}  
\caption{This shows the intensities of the transitions between a pair of asymmetry doublets as a function of Stark Field.   Parameters used for $J'' = K_{a}'' = M'' = 2 \rightarrow J' = K_{a}' = M' = 3$ transitions of propanediol}
\label{se-dn-1}
\end{center}
\end{figure}

\section{Effect of variation in excitation strength}

The above results were derived in the limit of weak excitation, we will now consider polarization pulses that move significant population between the two levels of a transition.    If $E_{\rm S}$ is sufficiently large that the individual Stark components $M'' \rightarrow M'$ are resolved in the excitation, then both $P_{X}$ and $P_{Y}$ FID signals will be proportional to $\sin( \theta)$ where $\theta = E_{\rm ex} \Delta t \left< J',K',M' |' \sum_{\alpha} \mu_{\alpha} \phi_{\alpha, X} | J'',K'',M'' \right>'/2\hbar$ is the ($E_{\rm S}$ dependent) excitation flip angle.  Thus, FID's of both polarizations will be optimized for  $\pi/2$ excitation.   In the high Stark field limit considered above, we will have a flip angle equals $|a|$ as given by Eq.~\ref{eqHFa}.    Note, however, that this is for a single  Stark component.  

If $E_{\rm S}$ is sufficiently weak that the transitions $\left| J'',\tau'',M''\right>  \rightarrow \left| J', \tau', M'' \pm 1\right>$ are both excited, then the pulse creates a coherent superposition of the $ \left| J', \tau', M'' \pm 1\right>$ states.   We first consider the Stark Field free case where we are pumping  $\left| J'',K'',M''\right> \rightarrow \left| J',K',M' = M'' \pm 1 \right>$ transitions.  Following an $X$ polarized excitation pulse we have
\begin{eqnarray}
& &\left| \psi(t,M'') \right> = \cos \left(\frac{ \omega^{0}(M'') \Delta t }{2}\right) \left| J'', \tau'', M'' \right> e^{-\frac{i E(J'',\tau'' )t}{ \hbar}} 
  + i \sin \left( \frac{ \omega^{0}(M'') \Delta t}{2} \right) e^{-\frac{i E(J',\tau' )t}{ \hbar}} \times \nonumber \\
& & \left(  \frac{\left| J', \tau', M'' + 1\right>\left<J', M''+1| \phi_{X}|J'',M'' \right> +  \left| J', \tau', M'' - 1\right>\left<J', M''-1 | \phi_{X}|J'',M''\right>  }{ \sqrt{|\left<J', M''+1| \phi_{X}|J'',M''\right>|^{2} + |\left<J', M''-1| \phi_{X}|J'',M''\right>|^{2}    }         }        \right)  \end{eqnarray}
with $M''$ dependent Rabi frequency given by:
\begin{eqnarray}
\omega^{0}(M'') &=& \frac{E_{\rm ex}}{\hbar}\left( \sum_{\alpha} \mu_{\alpha} \left< J',\tau' | \phi_{\alpha} | J'', \tau'' \right> \right) \nonumber \\
&& \sqrt{ |\left<J', M''+1| \phi_{X}|J'',M''\right>|^{2} + |\left<J', M''-1| \phi_{X}|J'',M''\right>|^{2}     } 
\end{eqnarray}
We can write 
\begin{eqnarray}
&& |\left<J', M''+1| \phi_{X}|J'',M''\right>|^{2} + |\left<J', M''-1| \phi_{X}|J'',M''\right>|^{2}   = \nonumber \\
&& \hspace{0.50in} 2(J''+1)(J''+2) + 2 M^{2}   \text{            if      } J' = J''+1 \nonumber \\
& &\hspace{0.50in}  2 J''(J''+1) - 2 M^{2}     \,\,\,\,\,\,\,\,\,\, \,\,\,\,\, \text{         if      }J' = J''
\end{eqnarray}

In the limit that we can treat the Stark Mixing with first order perturbation theory, we can write the Rabi frequencies linear in Stark Field as
\begin{eqnarray}
\omega(M'',E_{\rm S}) &=& \omega^{0}(M'') - E_{\rm S} \delta \omega(M'')  
\end{eqnarray}
\begin{eqnarray}
& &\delta \omega(M'') = \\
& & \sum_{\alpha,\beta} \frac{E_{\rm ex} \mu_{\alpha} \mu_{\beta}}{\hbar} \frac{ \left<J', M''+1| \phi_{X}|J'',M''\right> b_{+,\alpha,\beta}(M'')  +  \left<J', M''- 1| \phi_{X}|J'',M''\right> b_{-,\alpha,\beta}(M'')  }{\sqrt{ |\left<J', M''+1| \phi_{X}|J'',M''\right>|^{2} + |\left<J', M''-1| \phi_{X}|J'',M''\right>|^{2}      }} \nonumber 
\end{eqnarray}
\begin{eqnarray}
b_{\pm,\alpha,\beta}(M'') &=&  \sum_{J_{s}, \tau_{s}}   \left( \frac{ \left<  J', \tau', M'' \pm 1 \right| \phi_{\alpha,Z} \left| J_{s}, \tau_{s}, M'' \pm 1 \right>    
  \left<  J_{s}, \tau_{s}, M'' \pm 1 \right| \phi_{\beta,X} \left|  J'', \tau'', M''  \right>   }{E(J' ,\tau') - E(J_{s},\tau_{s})} \right. \nonumber \\ 
 &&  \hspace{0.3in} + \left.
  \frac{ \left<  J', \tau', M'' \pm 1 \right| \phi_{\beta,X} \left| J_{s}, \tau_{s}, M'' \right>    
  \left<  J_{s}, \tau_{s}, M'' \right| \phi_{\alpha,Z} \left|  J'', \tau'', M''  \right>   }{E(J'', \tau'') - E(J_{s},\tau_{s})} \right)
\end{eqnarray}
The polarization produced parallel to $Y$ will be:
\begin{eqnarray}
P_{Y} &=& \sum_{M'', \gamma} \frac{i}{2} \sin \left( \omega(M'',E_{\rm S}) \Delta t \right) \mu_{\gamma} \left< J'', \tau'', M'' | 
\phi_{\gamma, Y} \right| e^{-i \omega_{0}t} \times \\
& & \left(  \frac{\left| J', \tau', M'' + 1\right>\left<J', M''+1| \phi_{X}|J'',M'' \right> +  \left| J', \tau', M'' - 1\right>\left<J', M''-1 | \phi_{X}|J'',M''\right>  }{ \sqrt{ |\left<J', M''+1| \phi_{X}|J'',M''\right>|^{2} + |\left<J', M''-1| \phi_{X}|J'',M''\right>|^{2}      }         }        \right)
\nonumber \\
& &   + \text{c.c.} \nonumber \\
\nonumber
\end{eqnarray}
To first order in $E_{\rm S}$
\begin{equation}
\sin(\omega(M'',E_{\rm S}) \Delta t) = \sin \left(\omega^{0}(M'') \Delta t \right) - E_{\rm S}  \cos \left(\omega^{0}(M'') \Delta t \right) \delta \omega(M'') \Delta t
\end{equation}
As before, the term zero order in $E_{\rm S}$ generates no $Y$ polarized FID as long as the population of states with $M'' = +M$ and $-M$ are equal.  
\begin{eqnarray}
P_{Y} &=&  - \frac{i E_{\rm S}E_{\rm ex} \Delta t \mu_{a} \mu_{b} \mu_{c}}{2 \hbar} e^{-i \omega_{0}t} \sum_{M'', \alpha,\beta, \gamma} \cos \left( \omega^{0}(M'') \Delta t \right)  \left< J'', \tau'', M'' | 
\phi_{\gamma, Y} \right|  \nonumber \\
&\times & \left(  \frac{\left| J', \tau', M'' + 1\right>\left<J', M''+1| \phi_{X}|J'',M'' \right> +  \left| J', \tau', M'' - 1\right>\left<J', M''-1 | \phi_{X}|J'',M''\right>  }{ |\left<J', M''+1| \phi_{X}|J'',M''\right>|^{2} + |\left<J', M''-1| \phi_{X}|J'',M''\right>|^{2}               }        \right) \nonumber \\
&\times & \left( \left<J', M''+1| \phi_{X}|J'',M''\right> b_{+,\alpha,\beta}(M'')  +  \left<J', M''- 1| \phi_{X}|J'',M''\right> b_{-,\alpha,\beta}(M'')   
\right) + \text{c.c.} 
\end{eqnarray}

\section{Conclusions}
The breakthrough experiment of Patterson, Schnell \& Doyle~\cite{Patterson13} has demonstrated a novel way to study chiral molecules using microwave spectroscopy, which already allows (through isotopic substitution) the highest resolution way to determine equilibrium structure of isolated molecules but previously was not able to determine the chirality of a molecule.    This paper presents a thorough analysis of the theory behind this experiment and presents expressions that can be used to calculate the experimental observables in this and related experiments.   

\section{Acknowledgments}
The author would like to thank Brooks Pate for bringing Patterson, Schnell \& Doyle experiment to his attention and encouraging this project.


\newpage
\squeezetable
\begin{table}[htdp]
\caption{Calculated Chiral FID for $\kappa = -0.706  $}
\begin{center}
\begin{tabular}{|ccc|l|l|l|c|}
\hline \hline
J''	$_{	Kp''	,	Ko''	}$ & $\rightarrow$ &	J'	$_{	Kp'	,	Ko'	}$&	P(Y)	&	S	&	polarization	&	\\
\hline
0	$_{	0	,	0	}$ & $\rightarrow$ &	1	$_{	0	,	1	}$&	-0.014345093	&	0.333333333	&	a type	&	\\
0	$_{	0	,	0	}$ & $\rightarrow$ &	1	$_{	1	,	1	}$&	0.6651107	&	0.333333333	&	b type	&	\\
0	$_{	0	,	0	}$ & $\rightarrow$ &	1	$_{	1	,	0	}$&	-0.48409894	&	0.333333333	&	c type	&	\\
1	$_{	0	,	1	}$ & $\rightarrow$ &	2	$_{	0	,	2	}$&	-0.093857917	&	0.663570114	&	a type	&	\\
1	$_{	1	,	1	}$ & $\rightarrow$ &	2	$_{	0	,	2	}$&	-0.183313521	&	0.207471598	&	b type	&	\\
1	$_{	1	,	0	}$ & $\rightarrow$ &	2	$_{	0	,	2	}$&	0.129373065	&	0.128958289	&	c type	&	\\
1	$_{	0	,	1	}$ & $\rightarrow$ &	2	$_{	1	,	2	}$&	0.248745836	&	0.5	&	b type	&	\\
1	$_{	1	,	1	}$ & $\rightarrow$ &	2	$_{	1	,	2	}$&	0.176560314	&	0.5	&	a type	&	\\
1	$_{	0	,	1	}$ & $\rightarrow$ &	2	$_{	1	,	1	}$&	-0.103834802	&	0.5	&	c type	&	\\
1	$_{	1	,	0	}$ & $\rightarrow$ &	2	$_{	1	,	1	}$&	-0.104109539	&	0.5	&	a type	&	\\
1	$_{	1	,	1	}$ & $\rightarrow$ &	2	$_{	2	,	1	}$&	7.322764362	&	0.5	&	c type	&	\\
1	$_{	1	,	0	}$ & $\rightarrow$ &	2	$_{	2	,	1	}$&	7.042634499	&	0.5	&	b type	&	\\
1	$_{	1	,	1	}$ & $\rightarrow$ &	2	$_{	2	,	0	}$&	-6.442956114	&	0.459195069	&	b type	&	\\
1	$_{	1	,	0	}$ & $\rightarrow$ &	2	$_{	2	,	0	}$&	-7.832401565	&	0.537708378	&	c type	&	\\
2	$_{	0	,	2	}$ & $\rightarrow$ &	3	$_{	0	,	3	}$&	-0.312002831	&	0.988376378	&	a type	&	\\
2	$_{	1	,	2	}$ & $\rightarrow$ &	3	$_{	0	,	3	}$&	-0.170462573	&	0.464492753	&	b type	&	\\
2	$_{	1	,	1	}$ & $\rightarrow$ &	3	$_{	0	,	3	}$&	0.057202463	&	0.211955453	&	c type	&	\\
2	$_{	0	,	2	}$ & $\rightarrow$ &	3	$_{	1	,	3	}$&	0.230292826	&	0.69283623	&	b type	&	\\
2	$_{	1	,	2	}$ & $\rightarrow$ &	3	$_{	1	,	3	}$&	0.425185158	&	0.887727716	&	a type	&	\\
2	$_{	0	,	2	}$ & $\rightarrow$ &	3	$_{	1	,	2	}$&	-0.05625524	&	0.640647363	&	c type	&	\\
2	$_{	1	,	1	}$ & $\rightarrow$ &	3	$_{	1	,	2	}$&	-0.135165681	&	0.887417203	&	a type	&	\\
2	$_{	1	,	2	}$ & $\rightarrow$ &	3	$_{	2	,	2	}$&	1.051541456	&	0.555555556	&	c type	&	\\
2	$_{	1	,	1	}$ & $\rightarrow$ &	3	$_{	2	,	2	}$&	1.104838224	&	0.555555556	&	b type	&	\\
2	$_{	2	,	1	}$ & $\rightarrow$ &	3	$_{	2	,	2	}$&	0.010202148	&	0.555555556	&	a type	&	\\
2	$_{	1	,	2	}$ & $\rightarrow$ &	3	$_{	2	,	1	}$&	-0.841909744	&	0.424396136	&	b type	&	\\
2	$_{	1	,	1	}$ & $\rightarrow$ &	3	$_{	2	,	1	}$&	-1.249563689	&	0.676933435	&	c type	&	\\
2	$_{	2	,	0	}$ & $\rightarrow$ &	3	$_{	2	,	1	}$&	-0.015705437	&	0.555777842	&	a type	&	\\
2	$_{	2	,	1	}$ & $\rightarrow$ &	3	$_{	3	,	1	}$&	168.1415563	&	0.847860635	&	c type	&	\\
2	$_{	2	,	0	}$ & $\rightarrow$ &	3	$_{	3	,	1	}$&	152.5967058	&	0.816611135	&	b type	&	\\
2	$_{	2	,	1	}$ & $\rightarrow$ &	3	$_{	3	,	0	}$&	-152.1885753	&	0.81455008	&	b type	&	\\
2	$_{	2	,	0	}$ & $\rightarrow$ &	3	$_{	3	,	0	}$&	-168.4892139	&	0.849764516	&	c type	&	\\
\hline
\end{tabular}
\end{center}
\label{table:chi}
\end{table}

\begin{table}[htdp]
\caption{Calculated Chiral FID for $\kappa = -0.706  $, continued}
\begin{center}
\begin{tabular}{|ccc|l|l|l|c|}
\hline \hline
J''	$_{	Kp''	,	Ko''	}$ & $\rightarrow$ &	J'	$_{	Kp'	,	Ko'	}$&	P(Y)	&	S	&	polarization	&	\\
\hline
3	$_{	0	,	3	}$ & $\rightarrow$ &	4	$_{	0	,	4	}$&	-0.783932511	&	1.30828008	&	a type	&	\\
3	$_{	1	,	3	}$ & $\rightarrow$ &	4	$_{	0	,	4	}$&	-0.286248405	&	0.772842846	&	b type	&	\\
3	$_{	1	,	2	}$ & $\rightarrow$ &	4	$_{	0	,	4	}$&	0.035655368	&	0.245690525	&	c type	&	\\
3	$_{	0	,	3	}$ & $\rightarrow$ &	4	$_{	1	,	4	}$&	0.343979911	&	0.926718354	&	b type	&	\\
3	$_{	1	,	3	}$ & $\rightarrow$ &	4	$_{	1	,	4	}$&	0.892992397	&	1.246446683	&	a type	&	\\
3	$_{	0	,	3	}$ & $\rightarrow$ &	4	$_{	1	,	3	}$&	-0.047856102	&	0.738091198	&	c type	&	\\
3	$_{	1	,	2	}$ & $\rightarrow$ &	4	$_{	1	,	3	}$&	-0.125335263	&	1.24461383	&	a type	&	\\
3	$_{	2	,	2	}$ & $\rightarrow$ &	4	$_{	1	,	3	}$&	-0.257571118	&	0.187451354	&	b type	&	\\
3	$_{	2	,	1	}$ & $\rightarrow$ &	4	$_{	1	,	3	}$&	0.247715437	&	0.162755388	&	c type	&	\\
3	$_{	1	,	3	}$ & $\rightarrow$ &	4	$_{	2	,	3	}$&	0.278880711	&	0.613967307	&	c type	&	\\
3	$_{	1	,	2	}$ & $\rightarrow$ &	4	$_{	2	,	3	}$&	0.352860332	&	0.638828886	&	b type	&	\\
3	$_{	2	,	2	}$ & $\rightarrow$ &	4	$_{	2	,	3	}$&	0.035371194	&	0.998783251	&	a type	&	\\
3	$_{	1	,	3	}$ & $\rightarrow$ &	4	$_{	2	,	2	}$&	-0.214634099	&	0.3637627	&	b type	&	\\
3	$_{	1	,	2	}$ & $\rightarrow$ &	4	$_{	2	,	2	}$&	-0.367408049	&	0.866105215	&	c type	&	\\
3	$_{	2	,	1	}$ & $\rightarrow$ &	4	$_{	2	,	2	}$&	-0.048336179	&	1.000693648	&	a type	&	\\
3	$_{	2	,	2	}$ & $\rightarrow$ &	4	$_{	3	,	2	}$&	19.19218407	&	0.918062199	&	c type	&	\\
3	$_{	2	,	1	}$ & $\rightarrow$ &	4	$_{	3	,	2	}$&	16.89320368	&	0.823592697	&	b type	&	\\
3	$_{	3	,	1	}$ & $\rightarrow$ &	4	$_{	3	,	2	}$&	-0.000113317	&	0.584011457	&	a type	&	\\
3	$_{	2	,	2	}$ & $\rightarrow$ &	4	$_{	3	,	1	}$&	-16.65395456	&	0.812548646	&	b type	&	\\
3	$_{	2	,	1	}$ & $\rightarrow$ &	4	$_{	3	,	1	}$&	-19.38320717	&	0.928278022	&	c type	&	\\
3	$_{	3	,	0	}$ & $\rightarrow$ &	4	$_{	3	,	1	}$&	-0.001708194	&	0.584015534	&	a type	&	\\
3	$_{	3	,	1	}$ & $\rightarrow$ &	4	$_{	4	,	1	}$&	3806.704117	&	1.181293203	&	c type	&	\\
3	$_{	3	,	0	}$ & $\rightarrow$ &	4	$_{	4	,	1	}$&	3469.978829	&	1.150427611	&	b type	&	\\
3	$_{	3	,	1	}$ & $\rightarrow$ &	4	$_{	4	,	0	}$&	-3469.674136	&	1.150332728	&	b type	&	\\
3	$_{	3	,	0	}$ & $\rightarrow$ &	4	$_{	4	,	0	}$&	-3806.963474	&	1.181380843	&	c type	&\\
\hline
\end{tabular}
\end{center}
\label{table:chi2}
\end{table}

\end{document}